\documentclass{ws-p8-50x6-00}

\newcommand{\vek}[1]{{\mbox {\boldmath $#1$}}}

\begin{document}

\title{Neutralino Dark Matter and Caustic Ring Signals}

\author{Christofer Gunnarsson}

\address{Department of Physics, Stockholm University, Box 6730\\
SE-113 85 Stockholm, Sweden\\
E-mail: cg@physto.se}


\maketitle

\abstracts{We have studied the effects on the gamma-ray flux on 
Earth by neutralino annihilations in caustic rings of dark matter in 
the Galactic halo. The caustic ring model by Sikivie has been
used where dark matter particles are assumed to possess angular
momentum with respect to the Galactic centre. The computer code 
{\sffamily DarkSUSY} was
then used to examine the supersymmetric implications on the flux.  
We conclude that a small signal might be detectable under very optimistic
assumptions, but under these only.}   

\section{Introduction}
\label{sec:intro}
The overwhelming evidence for Dark Matter presents to us the intriguing 
questions of what it might consist of and how we can detect
it. Several WIMP (Weakly Interacting Massive Particle)
candidates have been suggested of which one is of particularly great interest,
the Lightest Supersymmetric Particle (LSP). In most scenarios this is the 
\emph{neutralino}. The possible properties of supersymmetric particles
can be examined 
with the \texttt{Fortran} computer code {\sffamily DarkSusy}\cite{darksusy} 
which has been used in this work, focusing on the 
(not yet discovered) neutralino. 

The experimental search for dark matter is an ongoing project employing many 
direct detection experiments, such as CDMS and DAMA. All show null results 
except for DAMA which claims an annual signal modulation due to the 
motion of the Earth about the sun and which could be explained by 
WIMPs\cite{dama}. Indirect detection experiments searching for MACHO
(Massive Astrophysical Compact Halo Object)-induced microlensing
events, such as MACHO and EROS conclude that MACHOs alone (of mass up
to 1 $M_{\odot}$) cannot make up the entire halo\cite{eros}.  

Another indirect detection possibility arises when studying the
properties of the neutralino. Being a Majorana particle, it can
self-annihilate and produce gamma photons. The annihilation rate is
proportional to the square of the particle density, implying that
wherever an overdensity is formed, an increased gamma flux will result
from that region. Overdensities may form in different ways and
recently through $N$-body simulations, Calc\'aneo-Rold\'an and Moore 
showed that primordial 
fluctuation-seeded hierarchical clustering of dark matter into clumps
is a most likely scenario\cite{calmor}. They also study the effects of
neutralino annihilation on the gamma ray flux on Earth. This work is
also supplemented in Ref.\cite{gunnarsson}. 

An almost orthogonal structure formation scenario is proposed by
Sikivie\cite{sik1}, where dark matter falls more or less smoothly onto
the Galaxy. The infall produces continuous regions of enhanced density, caustic
rings, in the dark matter halo of the Milky Way. This model and its
implications for the gamma flux is studied in greater detail in the
following sections.  
   
\section{Caustic rings of dark matter}
\subsection{Infall model}
\label{sec:infall}
Consider a set of particles initially positioned on a spherical
shell centered around an overdensity. If the age of the Universe is
small enough, their motion will be governed 
almost solely by the Hubble flow. Eventually they will be halted by the
gravity of the overdensity (galaxy), turn around, 
and start falling back onto the galaxy. They will then pass through
the centre of the galaxy, again reach a point of turnaround, 
again pass the galactic centre and so on. At turnaround, the particle
density will be enhanced since the phase-space sheet on which they  
lie will fold back there, they create a so called outer caustic. 

Assuming also that the particles have acquired angular momentum similar
to that of the luminous parts of the Galaxy, only the particles lying
on the axis of rotation will pass through the Galaxy centre. Particles
near the ``equator'' will produce another caustic near their point of
closest approach to the Galactic centre. These are called inner
caustics or caustic rings, since they will be circular density
enhancements. In fact, if the particles have vanishing initial
velocity dispersion, the density will diverge. The oscillation of
shells will be similar with the
presence of angular momentum and several caustic rings will be
produced with decreasing radii due to the deepening of the potential
well from the infall. Since the infall is continuous, the caustic
rings will be a persistent feature in space. It turns out that we seem
to be quite close to the fifth of these rings\cite{sik2}.

\subsection{Density profile}  
\label{denprof}
In the following we will neglect the velocity
dispersion when we derive the general shape and location of the
caustics, but introduce a small velocity dispersion 
when we consider the detailed density
distribution close to the caustics.

We label the particles arbitrarily by a
3-parameter, $\vek{\alpha}$ (which could, for instance, be the
position of the particle at a given initial time).  The flow of a
particle is completely specified by giving for each time its spatial
coordinate $\vek{\rm x}(\vek{\alpha},t)$.  If we have $n$ different flows
at $\vek{\rm x}$ and $t$, we can write the solutions of
$\vek{\rm x}=\vek{\rm x}(\vek{\alpha},t)$ as $\vek{\alpha}_{j}(\vek{\rm x},t)$,
where $j=1,\ldots ,n$.  To obtain the total number of particles, $N$,
we integrate the number density of particles over
$\vek{\alpha}$-space,
\begin{equation}
\label{eq:totnumpart}
N=\int\frac{d^3N\left(\vek{\alpha}\right)}{d\alpha_{1}d\alpha_{2}d\alpha_{3}}
d^{3}\alpha.
\end{equation}
Mapping onto position space gives the number density
\begin{equation}
\label{eq:numdens}
  d\left(\vek{\rm x},t\right)=\sum_{j=1}^{n}\frac{d^3N\left(\vek{\alpha}_{j}\left(\vek{\rm x},t\right)\right)}
{d\alpha_{1}d\alpha_{2}d\alpha_{3}}
\frac{1}{\left|D\left(\vek{\alpha},t\right)
  \right|_{\mbox{\scriptsize $\vek{\alpha}_{j}\left(\vek{\rm
x},t\right)$ }}},
\end{equation}
where $D\,$$\left(\vek{\alpha},t\right)$$\, \equiv\, $$\mathrm{det}$$\left(\frac{\partial{\mbox{\footnotesize{\vek{\rm x}}}}}
{\partial{\mbox{\footnotesize{\vek{\alpha}}}}}\right)$
is the Jacobian of the map $\vek{\alpha}\rightarrow \vek{\rm x}$.
Wherever $D\left(\vek{\alpha},t\right)=0$, the density will diverge, and hence caustic surfaces
are associated with zeros of $D$. 

We assume that the flow of particles is axially symmetric
about the $\hat{z}$-axis, coinciding with the rotation axis of the Galaxy
and also reflection symmetric with
respect to the $\hat{x}$-$\hat{y}$-plane, i.e.~under reflection
$z\rightarrow-z$. We also assume the dimensions of the cross-section
of the caustic ring to be small compared to the ring
radius. Cylindrical symmetry suggests using cylindrical coordinates.

\begin{figure}
\epsfxsize=18pc % will enlarge or reduce the postscript figures based on the xsize
\centerline{\epsfbox{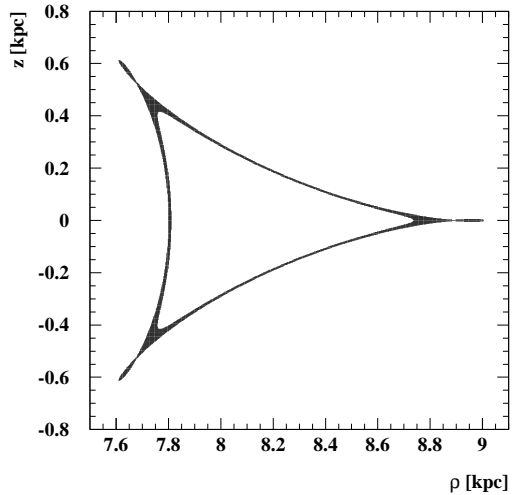}} % postscript image file name
\caption{Plot of points where the density exceeds 1.0 GeV/cm$^{3}$. Note that
  for ${\cal R}_{0}$ between about 7.8 and 8.8 kpc, we are situated inside
  the 'tricusp'. $\rho$ and $z$ are the distances from the Galactic centre 
  and plane respectively. }
\label{fig:tricuspquant}
\end{figure}

In Fig.~\ref{fig:tricuspquant} we show the cross section of the fifth
caustic ring (which is the one closest to us), where regions with a
density larger than 1 GeV/cm$^{3}$ have been indicated. 
We see that the caustic ring resembles a `tricusp'. 

The model of Sikivie has a set of caustic parameters that
describe it. 
To find these we assume that the turnaround sphere
is initially rigidly rotating and that it initially really is a sphere, not
just an axially symmetric topological sphere. We also have to make an 
assumption about 
the distribution of the smooth component of the dark matter 
distribution (i.e.~not associated with the caustic flows). We adopt
a time-independent logarithmic potential producing perfectly flat 
rotation curves. To obtain the caustic parameters,
we have followed the procedure in Ref.\cite{sik2}.  The interested
reader can find the result and more details about the caustic
parameters in Ref.\cite{cgexjobb}. 

A diverging density at the caustics results from our assumption of
zero velocity dispersion, which of course is an over-simplified
assumption.  We thus reintroduce a non-zero velocity dispersion by
estimating how much a given velocity dispersion would smear the
caustic.  We do that by considering a particle falling into the
aforementioned logarithmic potential.  
If we change the initial velocity of the particle
with the velocity dispersion, we obtain a difference in the location
of the point of closest approach (i.e.~the location of the caustic
ring).  We can then use this difference as an estimate of how much the
caustic ring is smeared by the velocity dispersion.  The simplest way
to take the smearing into account is to apply a cut-off in the density
whenever we are closer to the caustic than the smearing scale.  For a 
velocity dispersion of\cite{sik2} $\delta_{v}/c=10^{-12}$, 
this corresponds to a
cut-off in the density at ${\cal D}_{\rm cut} \simeq 250$
GeV/cm$^{3}$, which we used in the calculations.
To obtain the mass density, we multiply Eq.~(\ref{eq:numdens}) by the
particle mass, $m_{\chi}$, and the result can be found in
Fig.~\ref{fig:tricuspquant}, where the cross-section of the caustic
ring is plotted in cylindrical coordinates by showing all points 
where the density
exceeds 1 GeV/cm$^3$.  
\begin{figure}
\epsfxsize=20pc % will enlarge or reduce the postscript figures based on the xsize
\centerline{\epsfbox{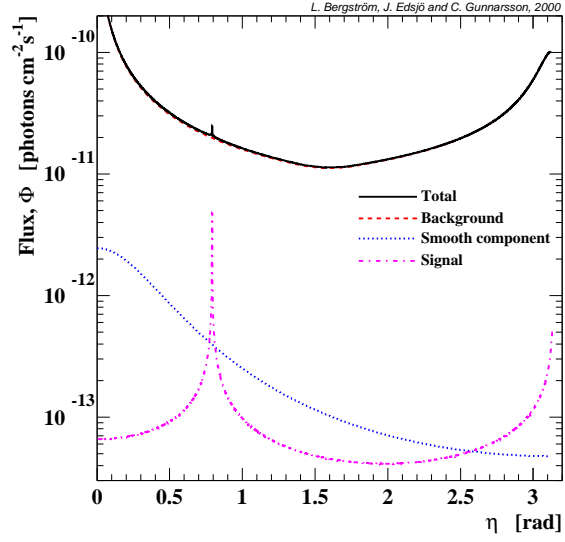}} % postscript image file name
\caption{Flux of gamma rays above 1 GeV from the caustic, the smooth
halo, the diffuse background and the sum of them for $\Delta\Omega=10^{-5}$
sr.  A maximum flux SUSY model was used for the signal and the smooth
part.  For the smooth halo a modified isothermal sphere with the scale radius
$a_{\rm c}=4$ kpc and the local density ${\cal D}_{0}=0.2$ GeV/cm$^{3}$
was used.  Our galactocentric distance was set to ${\cal R}_{0}=8.2$
kpc, but the results are essentially the same for other values.}
\label{fig:total}
\end{figure}

\section{Gamma-ray flux from neutralino annihilation}
\label{gamma}
Now assume that the dark matter in the caustics is in the form of
neutralinos. Self-annihilation in the overdense regions into gamma
rays will increase the $\gamma$-flux from that location. A continuous
gamma ray spectrum mainly coming from fragmentation of quark jets will
be the most prominent  
gamma annihilation channel. Monochromatic gammas are also possible but
will have a much smaller flux.

The $\gamma$-ray flux from WIMP annihilations in the galactic halo is
given by\cite{bergstromullio}
\begin{equation}
  \label{eq:flux1}
  \Phi_{\gamma}(\eta)=\frac{N_{\gamma}\sigma v}{4\pi
  m_{\chi}^{2}}\int_{L}{\cal D}^{2}(\ell )\ d\ell (\eta),
\end{equation}
where ${\cal D}(\ell )$ is the halo mass density of  WIMPs at distance $\ell$
along the line of sight, $\eta$ is the angle between the direction of
the Galactic centre and the line of sight, in the
$\hat{\rho}$-$\hat{z}$-plane. $N_{\gamma}$ is the number of photons
produced per annihilation. 

Assuming a supersymmetric model giving the largest flux, we show in
Fig.~\ref{fig:total} the expected flux from the caustic ring as a
function of $\eta$. We used a galactocentric distance of $\mathcal{R}_0=$8.2
kpc and Eq.~(\ref{eq:flux1}) was also integrated over the 
angular acceptance of a telescope of $\Delta\Omega=10^{-5}$ sr. 
Plotted in the figure is also a fit\cite{piero} to the EGRET data of the
diffuse background gamma flux, the expected flux from annihilations in the
smooth isothermal halo and the sum of the three. 
A significant peak can be seen in the directions of the cusps.  

\section{Conclusions}
As can be clearly seen in Fig.~\ref{fig:total}, there is a possibility for a
signal exceeding the diffuse background. However, this is under
very optimistic assumptions, such as a very small velocity dispersion 
and a maximum
flux supersymmetric model. Relaxing these would decrease the flux by
orders of magnitude and make the signal vanishingly small compared to
the background. 
If the optimistic assumptions unlikely turns out to be true, the Gamma
ray Large Area Space Telescope (GLAST) launched in 2005 would be able
to detect about 330 events from the closest parts of the caustic and about 
1700 events from the
diffuse background during one year in the same solid angle window of
$0.18^{\circ}\times 100^{\circ}$.
\raggedbottom
%\begin{figure}[t]
%\figurebox{20pc}{15pc}{} % to have a box alone
%\epsfxsize=10pc % will enlarge or reduce the postscript figures based on the xsize
%\epsfbox{xxx.eps} % postscript image file name
%\caption{ }
%\label{fig:radish}
%\end{figure}

\section*{Acknowledgments}
I would like to thank Joakim Edsj\"o and Lars Bergstr\"om for
invaluable help in the work and
I would also like to thank Pierre Sikivie for useful comments and
discussions on the topic.

\end{document}